\documentclass{aastex}
\usepackage{emulateapj5,times,mathptm}

\newcommand {\asec} {$^{\prime\prime}$}

\newcommand{\Lsolar}{\mbox{\,$\rm L_{\odot}$}}

\def\amin{\ifmmode ^{\prime}\else$^{\prime}$\fi}
\def\asec{\ifmmode ^{\prime\prime}\else$^{\prime\prime}$\fi}

\def\etal{{et\,al.\,}}

\def\chandra{{\it Chandra\/}}

\def\heao1{{\it HEAO-1\/}}

\def\iso{{\it ISOCAM\/}}

\def\ltsima{$\; \buildrel < \over \sim \;$}
\def\simlt{\lower.5ex\hbox{\ltsima}}
\def\gtsima{$\; \buildrel > \over \sim \;$}
\def\simgt{\lower.5ex\hbox{\gtsima}}

\begin{document}
%

%
\title{The Chandra Deep Field North Survey. XI. X-ray emission from luminous infrared starburst galaxies}
%

\author{D.M.~Alexander,\altaffilmark{1} H.~Aussel,\altaffilmark{2} F.E.~Bauer,\altaffilmark{1} W.N.~Brandt,\altaffilmark{1} A.E.~Hornschemeier,\altaffilmark{1} C.~Vignali,\altaffilmark{1} G.P.~Garmire,\altaffilmark{1} and D.P.~Schneider\altaffilmark{1}}

\affil{$^1$Department of Astronomy \& Astrophysics, 525 Davey Laboratory, The Pennsylvania State University, University Park, PA 16802}

\affil{$^2$Institute for Astronomy, 2680 Woodlawn Drive, Honolulu, HI 96822}


\shorttitle{X-RAY EMISSION FROM LUMINOUS INFRARED STARBURST GALAXIES}

\shortauthors{ALEXANDER ET AL.}

%
\begin{abstract}
%

Using the 1~Ms \chandra\ Deep Field North and 15~$\mu$m \iso\ HDF-N surveys we find a tight correlation between the population of strongly evolving starbursts discovered in faint 15~$\mu$m \iso\ surveys and the apparently normal galaxy population detected in deep X-ray surveys. Up to 100\% of the X-ray detected emission-line galaxies (ELGs) have 15~$\mu$m counterparts, in contrast to 10--20\% of the X-ray detected absorption-line galaxies and AGN-dominated sources. None of the X-ray detected ELGs is detected in the hard band (2--8~keV), and their stacked-average X-ray spectral slope of $\Gamma\approx2.0$ suggests a low fraction of obscured AGN activity within the X-ray detected ELG population. The characteristics of the $z=$~0.4--1.3 X-ray detected ELGs are consistent with those expected for M82 and NGC~3256-type starbursts; these X-ray detected ELGs contribute $\approx$~2\% of the 0.5--8.0~keV X-ray background. The only statistical difference between the X-ray detected and X-ray undetected 15~$\mu$m selected ELGs is that a much larger fraction of the former have radio emission.

\end{abstract}


%
\section{Introduction}\label{intro}
%

Faint 15~$\mu$m \iso\ surveys have revealed a large population of strongly evolving luminous infrared (IR) starbursts at $z\approx1$ (e.g.,\ Aussel \etal 1999; Elbaz \etal 1999; Elbaz \etal 2002). The space density of these sources is an order of magnitude higher than that found for luminous IR starbursts in the local Universe (i.e.,\ $z\simlt0.1$). Although these sources only account for a small fraction of the star-forming galaxy population, they are likely to contribute a significant fraction of the star-formation history of the Universe (e.g., Chary \& Elbaz 2001; Elbaz \etal 2002).

In this Letter we use the 1~Ms \chandra\ Deep Field-North (CDF-N; Brandt \etal 2001b, hereafter Paper V) and \iso\ HDF-N (Aussel \etal 1999, 2002) surveys to show that these strongly evolving luminous IR starbursts are identified with the population of apparently normal galaxies detected in deep X-ray surveys (e.g.,\ Giacconi \etal 2001; Hornschemeier \etal 2001, hereafter Paper II; Brandt \etal 2001a, hereafter Paper IV). Previous cross-identification studies have shown that $\approx$~20\% of the \iso\ sources are AGN-dominated; however, few constraints have been placed on the larger fraction of \iso\ sources that do not have obvious AGN activity (e.g.,\ Paper II; Paper IV; Fadda \etal 2002). The primary focus of this paper is to place constraints on the X-ray properties of these sources. The Galactic column density toward the CDF-N is $(1.6\pm 0.4)\times 10^{20}$~cm$^{-2}$ (Stark \etal 1992), and $H_{0}=65$~km~s$^{-1}$~Mpc$^{-1}$, $\Omega_{\rm M}=\onethird$, and $\Omega_{\Lambda}=\twothirds$ are adopted throughout this Letter.

%
\section{Infrared and X-ray observations}\label{data}
%

The 1~Ms CDF-N observations were centered on the HDF-N (Williams \etal 1996) and cover $\approx450$~arcmin$^2$ (Paper~V). The $\approx$~21.5~arcmin$^2$ region of the CDF-N survey coincident with the 15~$\mu$m \iso\ observations is close to the CDF-N aim point. Forty-nine high-significance X-ray sources (i.e.,\ {\sc wavdetect} false-positive probability threshold of 10$^{-7}$) are detected in this area down to on-axis 0.5--2.0~keV (soft-band) and 2--8~keV (hard-band) flux limits of $\approx3\times10^{-17}$~erg~cm$^{-2}$~s$^{-1}$ and $\approx2\times10^{-16}$~erg~cm$^{-2}$~s$^{-1}$, respectively (Paper V).

The \iso\ observations were performed at 6.7~$\mu$m and 15~$\mu$m. Because of the significantly smaller field-of-view of the 6.7~$\mu$m observations and the small number of sources detected, only the 15~$\mu$m observations are considered here. Discrete sources are detected down to $f_{\rm 15\mu m}\approx$~20~$\mu$Jy, although the \iso\ observations are far from complete at this depth. In this study we focus on the 41 sources with $f_{\rm 15\mu m}\ge$~100~$\mu$Jy in the complete 15~$\mu$m selected sample of Aussel \etal (2002).

%
\section{The Infrared-X-ray Connection}\label{irx}
%

\subsection{Basic source properties}

Optical ($I$-band) sources taken from the photometric catalog produced in Alexander \etal (2001b, hereafter Paper VI) were matched to X-ray and 15~$\mu$m sources using matching radii of 1\asec\ and 3\asec, respectively (see Paper VI; Aussel \etal 2002).\footnote{The $I$-band photometric catalog was produced using the Barger \etal (1999) image publicly available from http://www.ifa.hawaii.edu/$\sim$cowie/hdflank/hdflank.html.} The 15~$\mu$m sources have $I=$~18--23, and all but one source have redshifts in the catalogs of Cohen \etal (2000), Cohen (2001) or Dawson \etal (2001). By contrast, $\approx$~40\% of the X-ray sources in the 15~$\mu$m \iso\ region have $I\ge23$, although all but two of the $I<23$ sources have redshifts. 

In total 14 of the 41 15~$\mu$m sources have X-ray counterparts in the high-significance X-ray source catalog.\footnote{X-ray sources detected in the soft band (0.5--2.0~keV), hard band (2--8~~keV) or full band (0.5--8.0~keV) are matched to 15~$\mu$m counterparts.} However, given the comparitively low source density of 15~$\mu$m sources, we can search for lower significance X-ray sources associated with 15~$\mu$m sources without introducing a significant number of spurious X-ray detections. We ran {\sc wavdetect} with a false-positive probability threshold of 10$^{-5}$ and found six further 
%
%
\vspace{0.2in}
\centerline{\includegraphics[angle=0,width=9.5cm]{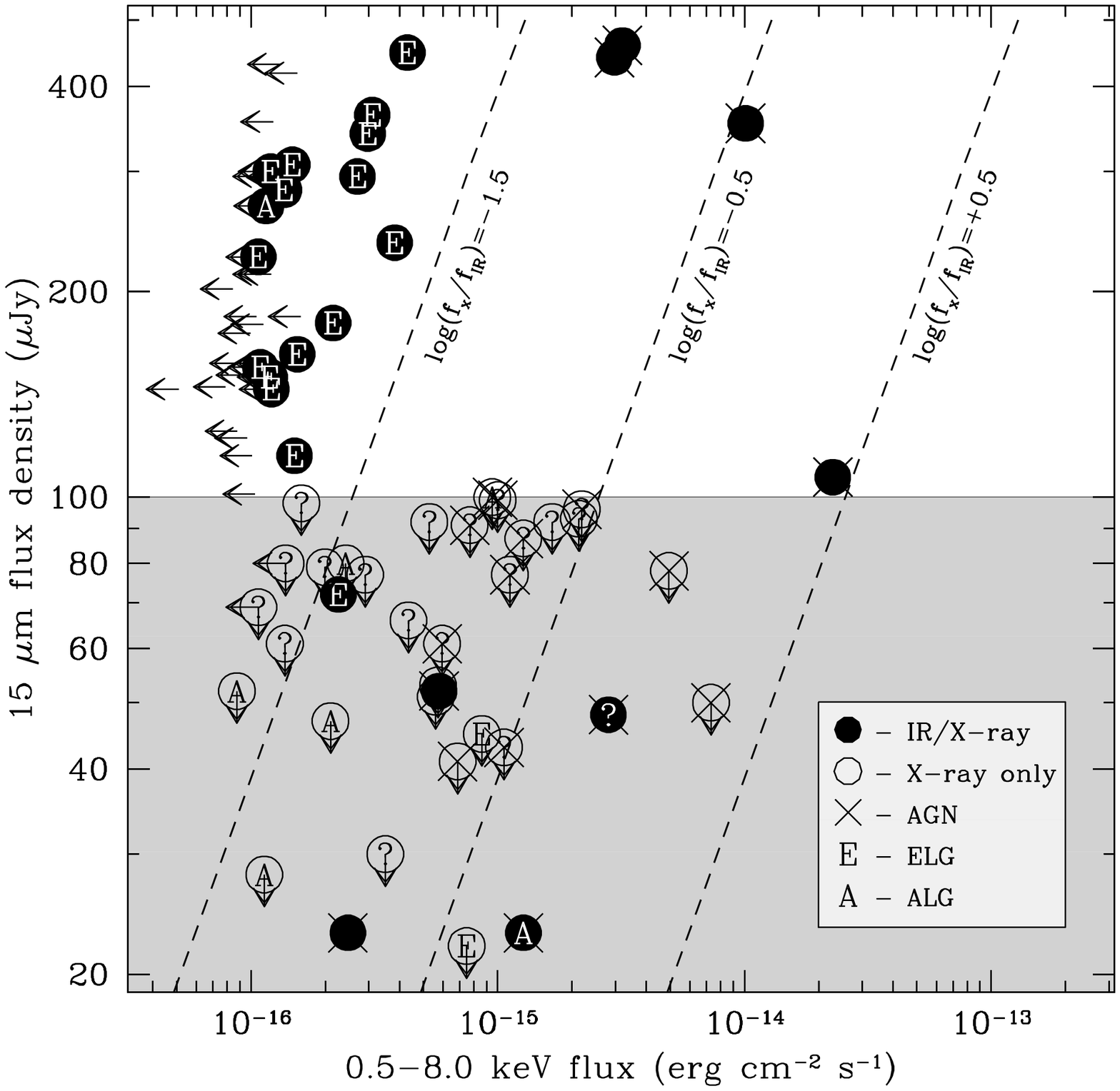}}
\vspace{-0.1in}
\figcaption{Full-band X-ray flux versus 15~$\mu$m flux density. The filled circles are the X-ray-IR matched sources, and the open circles are the X-ray-only sources. Characters inside the circles indicate different source types: ``E'' indicates ELGs, ``A'' indicates ALGs, ``?'' indicates sources that do not have spectroscopic classifications, and overlaid crosses indicate AGN-dominated sources; see \S3.2. IR sources not detected with X-ray emission are plotted only as upper limit arrows. The diagonal lines indicate constant flux ratios and the unshaded region indicates the IR sources that lie within the 15~$\mu$m sample definition.}
\label{fig:redshift1}
\vspace{0.4in}


\noindent matches to 15~$\mu$m sources; we would expect $\approx$~0.01 of these six X-ray sources to be spurious. Finally, we calculated X-ray upper limits for the 21 X-ray undetected 15~$\mu$m sources following \S3.2.1 of Paper V. The fraction of 15~$\mu$m sources with X-ray counterparts is $49^{+14}_{-11}$\% (see Figure~1).\footnote{All errors are taken from Tables 1 and 2 of Gehrels (1986) and correspond to the $1\sigma$ level; these were calculated assuming Poisson statistics.}

We also matched the X-ray sources to the 15~$\mu$m sources in the fainter 15~$\mu$m sample of Aussel \etal (1999) and computed 15~$\mu$m 3$\sigma$ upper limits for all the X-ray sources without 15~$\mu$m counterparts following Aussel \etal (2002). Five X-ray sources have 15~$\mu$m detected counterparts with $f_{\rm 15\mu m}<$~100~$\mu$Jy (see Figure~1); these 15~$\mu$m sources fall below our sample threshold and are not included in any further discussion.

\subsection{Source-type classification}

We have classified our sources since AGN-dominated sources are predicted to have different X-ray-to-IR flux ratios than non-AGN dominated sources (see Barcons \etal 1995; Alexander \etal 2001a). We adopt three source-type classifications: AGN-dominated sources, emission-line galaxies (ELGs), and absorption-line galaxies (ALGs). We considered a source to be AGN-dominated if it has either a ``Q'' (i.e.,\ broad emission lines) optical spectral classification from Cohen \etal (2000), a rest-frame 0.5--8.0~keV 
%
%
\vspace{0.2in}
\centerline{\includegraphics[angle=0,width=9.5cm]{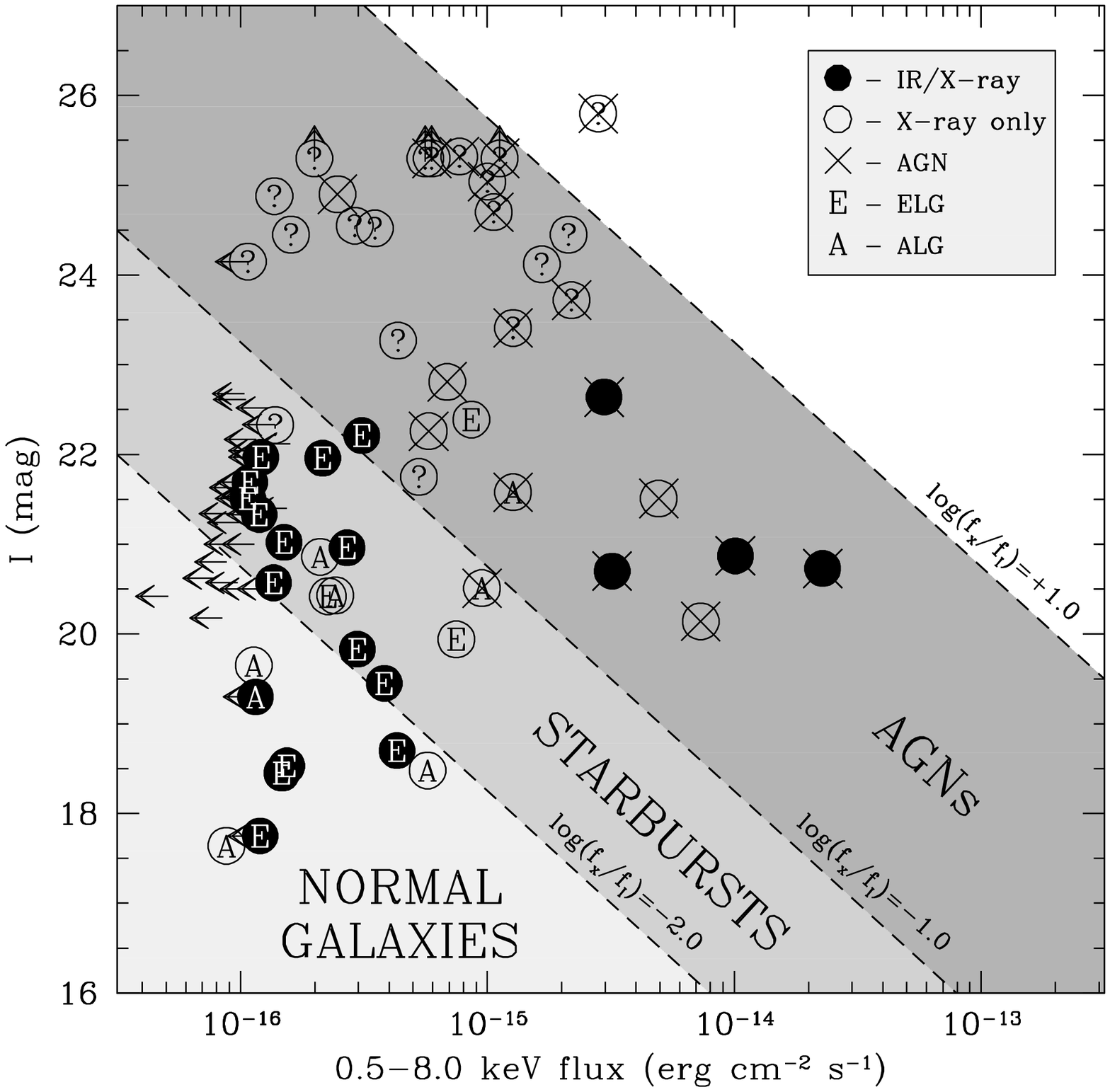}}
\vspace{-0.1in}
\figcaption{Full-band X-ray flux versus $I$-band magnitude. All symbols have the same meaning as in Figure~1. The diagonal lines indicate constant flux ratios. The shaded regions show the flux ratios of the AGN-dominated sources, X-ray detected starbursts and X-ray detected normal galaxies (see \S3.3 and \S4 for justification); these regions should be considered approximate.}
\label{fig:redshift1}
\vspace{0.4in}


\noindent luminosity $>3\times10^{42}$ erg~s$^{-1}$, or a flat effective X-ray spectral slope (i.e.,\ $\Gamma<1.0$, an indicator of obscured AGN activity).\footnote{All X-ray luminosities in this paper are calculated conservatively assuming $\Gamma=2.0$ and no intrinsic or Galactic absorption.}$^{,}$\footnote{When determining if a source has a flat X-ray spectral slope, we only considered sources bright enough to place meaningful constraints (see \S3.2.1 in Paper V).} Clearly there can be AGN-dominated sources not selected by these criteria (e.g.,\ an AGN with $\Gamma>1.0$ and a rest-frame 0.5--8.0~keV luminosity $<3\times10^{42}$ erg~s$^{-1}$). We considered a source to be an ELG if it is not AGN dominated and has a classification including an ``E'' or ``I'' in Cohen \etal (2000). All sources with an ``A'' classification in Cohen \etal (2000) are considered ALGs; ALGs comprise both elliptical and quiescent spiral galaxies and can be AGN dominated.

\subsection{AGN-dominated sources}

In Figure~2 we show the full-band X-ray flux versus $I$-band magnitude for all of the 15~$\mu$m and X-ray detected sources. Almost all of the 19 AGN-dominated sources have $\log{({{f_{\rm X}}\over{f_{\rm I}}})}>-1$, consistent with the findings from previous X-ray surveys (e.g.,\ Schmidt \etal 1998; Akiyama \etal 2000). Four AGN-dominated sources have 15~$\mu$m counterparts. The resulting fraction ($21^{+18}_{-11}$\%) is consistent with that found by Fadda \etal (2002) and should be considered an upper limit since many of the unclassified sources with $\log{({{f_{\rm X}}\over{f_{\rm I}}})}>-1$ are probably also AGN dominated. These four AGN-dominated sources are among the X-ray brightest sources (i.e.,\ full-band fluxes $>2\times10^{-15}$~erg~cm$^{-2}$~s$^{-1}$). Assuming the highest X-ray-to-IR flux ratio of these AGN-dominated sources [i.e., $\log{({{f_{\rm X}}\over{f_{\rm IR}}})}\approx$~0.5], an AGN-dominated source at the X-ray flux limit of our survey will have $f_{\rm 15\mu m}\approx$~1~$\mu$Jy; see \S4 for discussion of deeper IR observations.

\subsection{Emission-line galaxies}

Fifteen of the 18 X-ray detected ELGs have 15~$\mu$m counterparts; see Figure~2. The three sources not detected at 15~$\mu$m have distinctive properties suggesting that their X-ray emission may well be AGN dominated. For example, all three sources have hard-band counterparts while none of the 15 sources detected at 15~$\mu$m has a hard-band counterpart. Furthermore, one of the three sources has an X-ray-to-optical flux ratio typical of an AGN-dominated source while another source is only detected in the hard band, suggesting it contains an obscured AGN.\footnote{This latter source was not bright enough to place meaningful constraints on its X-ray spectral slope (see \S3.2.1 in Paper V).} These three 15~$\mu$m undetected sources are excluded from further analysis of the X-ray detected ELGs as they are probably AGN dominated. Therefore, our results suggest that up to 100\% of the X-ray detected ELGs (without AGN) have 15~$\mu$m counterparts. Since 15~$\mu$m detected ELGs only account for $\approx$~20\% of the $R<23.5$ ELG population in this region [as determined from the Cohen \etal (2000) data], this implies an association between the production of X-ray and IR emission in these sources.

All of the X-ray detected ELGs have faint X-ray emission (i.e.,\ full-band fluxes $<5\times10^{-16}$~erg~cm$^{-2}$~s$^{-1}$), showing clear differences in the X-ray-to-IR and X-ray-to-optical emission relationships of AGN-dominated sources and X-ray detected ELGs (see Figures~1 and 2). While these sources are clearly not AGN-dominated, a fraction may contain a hidden (i.e.,\ obscured) AGN as revealed by a flat X-ray spectral slope. Although none of the sources is detected in the hard band, we can stack the individual non-detections to provide a statistical constraint following \S3.3 of Paper VI. This provides a detection in the hard band corresponding to an average band ratio (the ratio of hard-band to soft-band count rate) of $0.21\pm0.04$, corresponding to an effective photon index of $\Gamma\approx$~2.0. To help interpret this result, we also stacked the 17 optically faint (i.e.,\ $I\ge24$) X-ray sources in Paper VI with full-band fluxes $<5\times10^{-16}$~erg~cm$^{-2}$~s$^{-1}$; the majority of these sources are thought to be obscured AGN. The average band ratio from this stacking analysis is $0.62^{+0.09}_{-0.08}$ ($\Gamma\approx$~1.3), substantially higher than that found for the ELGs. The reason for this difference is clear: $59^{+25}_{-18}$\% of the optically faint X-ray sources have hard-band detections. Since we applied the same full-band flux limit to both samples, and thus took into account the high hard-band background, these results suggest a low fraction of obscured AGN within the ELG sample.

\subsection{Absorption-line galaxies}

Only one ($13^{+30}_{-11}$\%) of the eight X-ray detected ALGs has a 15~$\mu$m counterpart, in stark contrast to the higher 15~$\mu$m matching fraction found for the X-ray detected ELGs; see Figure~2. This difference could be due to either enhanced X-ray emission (e.g.,\ an AGN component) or decreased IR emission (e.g.,\ a weaker dust-emission component). Indeed, two ALGs are classified as AGN-dominated (see Figure~2), and the radio properties of a further fraction also indicate AGN activity (Bauer \etal 2002). However, if the absence of 15~$\mu$m counterparts in the other sources is due to a weaker dust-emission component, then their 15~$\mu$m emission is likely to be dominated by 
%
%
\vspace{0.2in}
\centerline{\includegraphics[angle=0,width=9.5cm]{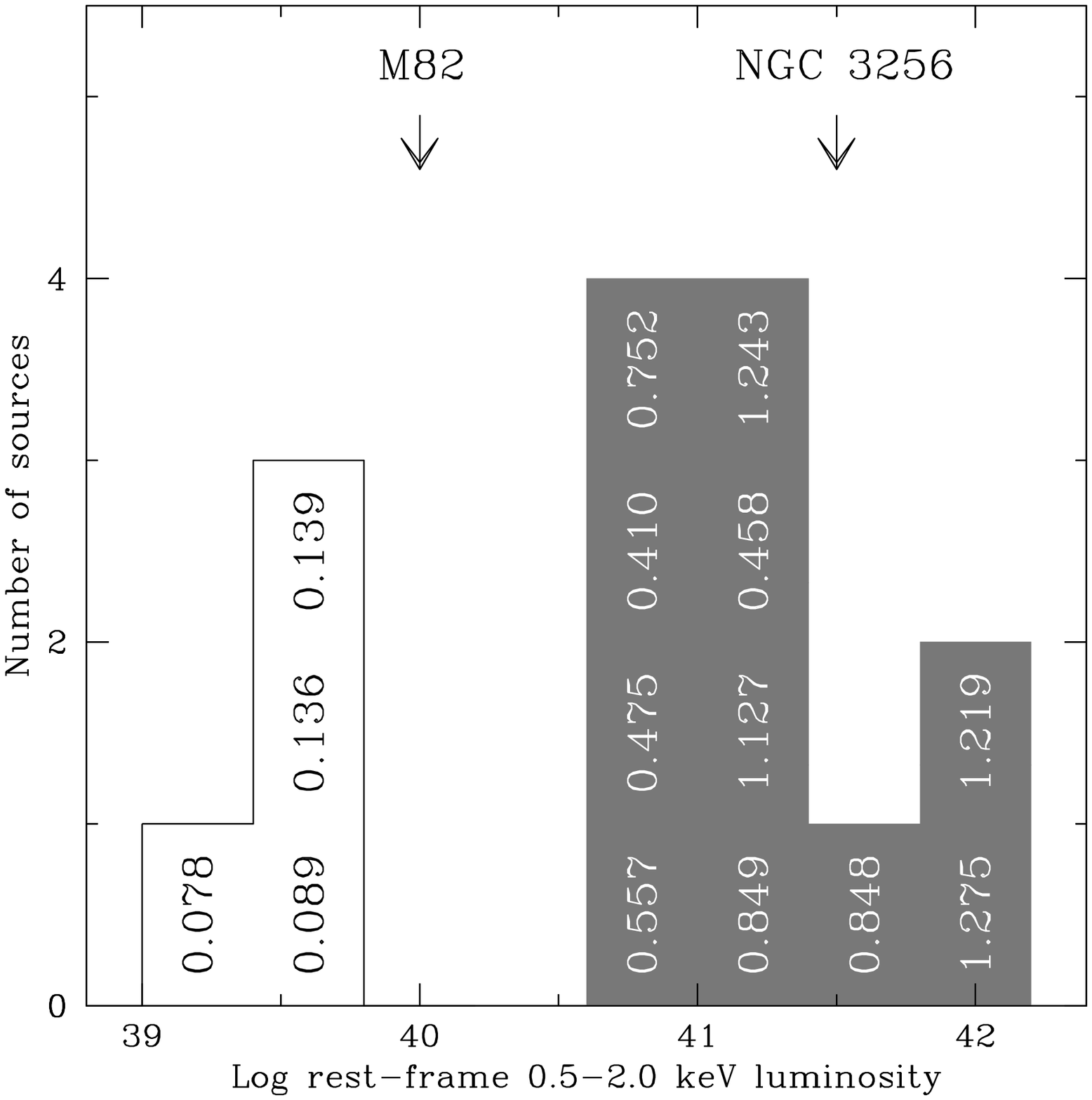}}
\vspace{-0.1in}
\figcaption{Rest-frame soft-band X-ray luminosity histogram for the X-ray detected ELGs with 15~$\mu$m counterparts. Individual source redshifts are shown (compare to Figure~2 in Paper VIII). The soft-band X-ray luminosities of the starburst galaxies M82 and NGC~3256 are indicated. The apparently bi-modality in the X-ray luminosities of the normal galaxies (unshaded) and starburst galaxies (shaded) is due to the small number of objects investigated here.}
\label{fig:redshift1}
\vspace{0.4in}


\noindent starlight. Following \S3.4.3 of Alexander \etal (2002), the estimated 15~$\mu$m flux densities from starlight for these sources are 5--40~$\mu$Jy, below the flux threshold of the 15~$\mu$m sample. Direct evidence for starlight emission is found in one source that is detected at 6.7~$\mu$m but not at 15~$\mu$m (CXOHDFN~J123649.5+621345; see Paper IV). Our estimated 15~$\mu$m flux density ($f_{\rm 15\mu m}=$~40~$\mu$Jy) is consistent with the upper limit for this source ($f_{\rm 15\mu m}<$~52~$\mu$Jy). 

%
\section{Discussion}
%

Studies of the faint 15~$\mu$m source population have shown that these sources are mostly luminous IR starburst galaxies undergoing intense dust-enshrouded star-formation activity that is re-radiated at IR wavelengths (e.g.,\ Chary \& Elbaz 2001). By cross-identifying the sources detected in the 1~Ms CDF-N and 15~$\mu$m \iso\ HDF-N surveys, we have shown that this starburst population is associated with the population of apparently normal galaxies detected at faint X-ray fluxes (e.g.,\ Giacconi \etal 2001; Paper II; Paper VI). With small X-ray-to-optical and X-ray-to-IR flux ratios [i.e.,\ $\log{({{f_{\rm X}}\over{f_{\rm I}}})}\simlt-1$ and $\log{({{f_{\rm X}}\over{f_{\rm IR}}})}<-1.5$, respectively], these X-ray detected ELGs are clearly different from AGN-dominated sources. None of the X-ray detected ELGs is detected in the hard band (2--8~keV), and their stacked-average X-ray spectral slope of $\Gamma\approx2.0$ suggests a low fraction of obscured AGN activity within the X-ray detected ELG population.

The range of rest-frame soft-band luminosities for the 11 $z=$~0.4--1.3 X-ray detected ELGs suggests they are probably X-ray detected starburst galaxies similar to M82 and NGC~3256 (see Figure~3; e.g.,\ Griffiths \etal 2000; Lira \etal 2002). The most X-ray luminous sources are up to $\approx$~3 times more luminous than NGC~3256, the most X-ray luminous nearby starburst galaxy. The range of $\nu$$L_{15\mu m}$ luminosities ($\approx2\times10^{10}$ to $\approx2\times10^{12}$\Lsolar\, depending on the $K$-correction used; e.g.,\ Elbaz \etal 2002) are also consistent with M82 and NGC~3256-type starbursts.\footnote{We note that far-IR background and sub-mm source-count constraints restrict the number of possible Arp~220-type galaxies (e.g.,\ Chary \& Elbaz 2001).} Moran, Lehnert, \& Helfand (1999) predicted that X-ray detected starbursts would contribute $\approx$~5--30\% of the X-ray background (XRB). However, the total soft and hard band fluxes of the X-ray detected starbursts in this study ($7.7\times10^{-16}$~erg~cm$^{-2}$~s$^{-1}$ and $1.5\times10^{-15}$~erg~cm$^{-2}$~s$^{-1}$, respectively) only account for $\approx$~2\% of the XRB (assuming the XRB flux densities given in Garmire \etal 2002). This fraction is below the lower limit given by Moran \etal (1999), although we note that these X-ray observations have probably not fully resolved the XRB (e.g.,\ Cowie \etal 2002) and we might expect these sources to dominate the faint X-ray source counts.

The large fraction of X-ray detected ELGs with 15~$\mu$m counterparts suggests an association between the production of X-ray and IR emission. Since the 15~$\mu$m emission is considered an unobscured indicator of star-formation rate (e.g.,\ Rowan-Robinson \etal 1997), this further suggests that the X-ray emission can provide an indication of star-formation rate. However, we urge caution since the 17 15~$\mu$m detected ELGs undetected at X-ray energies have $I$-band magnitude, $I-K$ color, $V-I$ color, 15~$\mu$m flux density, and redshift distributions indistinguishable from the 11 X-ray detected starbursts (a Kolmogorov-Smirnov test gives probabilities of 98\%, 97\%, 48\%, 78\% and 55\%, respectively).\footnote{The $I-K$ and $V-I$ colors were taken from the photometric catalog produced in Paper VI.} These results suggest a broad range of X-ray luminosities for starbursts of similar IR luminosities and implies that there may be another factor that effects the production of the X-ray emission. The lack of a statistical difference between the $V-I$ and $I-K$ colors suggest that galaxy type and dust extinction effects are not important. However, we note that a much larger fraction of the X-ray detected starbursts have radio counterparts; see Bauer \etal (2002) for further constraints.

The four $z<0.2$ sources have X-ray luminosities typical of normal galaxies (i.e.,\ $\simlt10^{40}$~erg~s$^{-1}$; see Figure~3; Hornschemeier \etal 2002, hereafter Paper VIII). Their average $\nu$$L_{15\mu m}$ luminosity of $\approx6\times10^{8}$\Lsolar\ is also typical of normal galaxies. These sources arise at smaller X-ray-to-optical flux ratios than the X-ray detected starbursts [i.e.,\ $\log{({{f_{\rm X}}\over{f_{\rm I}}})}\simlt-2$]. Thus to detect higher redshift normal galaxies will require either deeper \chandra\ observations or the application of X-ray stacking techniques (e.g.,\ Paper VIII).

Deeper IR observations (i.e.,\ GOODS\footnote{For details on GOODS, see http://www.stsci.edu/science/goods/.}) are planned with {\it SIRTF} over $\approx$~160~arcmin$^2$ of the most sensitive region of the CDF-N: these should reach a depth in the 24~$\mu$m MIPS band to an equivalent 15~$\mu$m flux density of $\approx$~10~$\mu$Jy. In addition to providing a sample of $\approx$~100 X-ray detected ELGs with 24~$\mu$m counterparts, these observations should be deep enough to detect counterparts for the majority of the X-ray detected ALGs and AGN-dominated sources.

%
\section*{Acknowledgments}
%

This work would not have been possible without the support of the entire \chandra\ and ACIS teams.
We acknowledge the financial support of
NASA grant NAS~8-38252 (GPG, PI),
NSF CAREER award AST-9983783 (DMA, FEB, WNB, CV),  
NASA GSRP grant NGT5-50247 and 
the Pennsylvania Space Grant Consortium (AEH), and
NSF grant AST-9900703~(DPS).

%

%
%



\begin{references}
%


\reference{}
Akiyama, M., et~al. 2000, ApJ, 532, 700

\reference{}
Alexander, D.M., et~al. 2001a, ApJ, 554, 18

\reference{}
Alexander, D.M., Brandt, W.N., Hornschemeier, A.E., Garmire, G.P., Schneider, D.P., Bauer, F.E., \& Griffiths, R.E. 2001b, AJ, 122, 2156 (Paper VI)

\reference{}
Alexander, D.M., Vignali, C., Bauer, F.E., Brandt, W.N., Hornschemeier, A.E., Garmire, G.P., \& Schneider, D.P. 2002, AJ, 123, 1149




\reference{}
Aussel, H., Cesarsky, C.J., Elbaz, D., \& Starck, J.L. 1999, A\&A, 342, 313

\reference{}
Aussel, H., et~al. 2002, in preparation

\reference{}
Barcons, X., Franceschini, A., De Zotti, G., Danese, L., \& Miyaji, T. 1995, ApJ, 455, 480

\reference{}
Barger, A.J., Cowie, L.L., Trentham, N., Fulton, E., Hu, E.M., Songaila, A., \& Hall, D.\ 1999, AJ, 117, 2656




\reference{}
Bauer, F.E., et~al. 2002, in preparation




\reference{}
Brandt, W.N., et~al. 2001a, AJ, 122, 1 (Paper IV)

\reference{}
Brandt, W.N., et~al. 2001b, AJ, 122, 2810 (Paper V)



\reference{}
Chary, R.~\& Elbaz, D.\ 2001, \apj, 556, 562



\reference{}
Cohen, J.G., Hogg, D.W., Blandford, R., Cowie, L.L., Hu, E., Songaila, A., Shopbell, P., \& Richberg, K. 2000, ApJ, 538, 29  

\reference{}
Cohen, J.G. 2001, AJ, 121, 2895





\reference{}
Cowie, L.L., Garmire, G.P., Bautz, M.W., Barger, A.J., Brandt, W.N., \& Hornschemeier, A.E. 2002, ApJ, 566, L5




\reference{}
Dawson, S., Stern, D., Bunker, A.J., Spinrod, H. \& Dey, A. 2001, AJ, 122, 598







\reference{}
Elbaz, D., Cesarsky, C.J., Chanial, P., Aussel, H., Franceschini, A., Fadda, D., \& Chary, R.R. 2002, A\&A, in press (astro-ph/0201328)


\reference{}
Fadda, D., Flores, H., Hasinger, G., Franceschini, A., Altieri, B., Cesarsky, C., Elbaz, D., \& Ferrando, P. 2002, A\&A, in press (astro-ph/0111412)





\reference{}
Garmire, G.P., et~al. 2002, ApJ, submitted

\reference{}
Gehrels, N. 1986, ApJ, 303, 336


\reference{}
Giacconi, R., et~al. 2001, ApJ, 551, 624


\reference{}
Griffiths, R.~E., Ptak, A., Feigelson, E.~D., Garmire, G., Townsley, L., Brandt, W.~N., Sambruna, R., \& Bregman, J.~N.\ 2000, Science, 290, 1325





\reference{}
Hornschemeier, A.E., et~al. 2001, ApJ, 554, 742 (Paper II)

\reference{}
Hornschemeier, A.E., et~al. 2002, ApJ, in press (astro-ph/0110094; Paper VIII)









\reference{}
Lira, P., Ward, M., Zezas, A., Alonso-Herrero, A., \& Ueno, S. 2002, MNRAS, 330, 259






\reference{}
Moran, E.C., Lehnert, M.D., \& Helfand, D.J. 1999, ApJ, 526, 649















\reference{}
Rowan-Robinson, M.~et al.\ 1997, \mnras, 289, 490




\reference{}
Schmidt, M., et~al. 1998, \aap, 329, 495






\reference{}
Stark, A.A., Gammie, C.F., Wilson, R.W., Bally, J., Linke, R.A., Heiles, C., \& Hurwitz, M. 1992, ApJS, 79, 77 



\reference{}
Williams, R.E., et~al.\ 1996, AJ, 112, 1335

\end{references}
\end{document}